\documentclass[twocolumn,preprintnumbers,amsmath,amssymb]{revtex4}

\usepackage{amsmath}    % need for subequations
\usepackage{graphicx}   % for figures
\usepackage{color}

\begin{document}

\title{Quantum fields are not fields. Comment on ``There are no particles, there are only fields,'' by Art Hobson [Am. J. Phys. 81(3), 211--223 (2013)]}
\author{Massimiliano Sassoli de Bianchi}
\affiliation{Laboratorio di Autoricerca di Base, 6914 Carona, Switzerland}\date{\today}
\email{autoricerca@gmail.com}   %optional

\maketitle

In Ref.~\onlinecite{Hob}, Hobson defends the superiority of the field concept in quantum physics, in comparison to the particle concept. His view is that if we acknowledge that the fundamental constituents of physical reality are fields, and not particles, then much of the interpretational difficulties of quantum physics would disappear. However, as we shall briefly explain in the present comment, quantum fields are no more fields than quantum particles are particles, so that the replacement of a particle ontology (or particle and field ontology) by an all-field ontology, will not solve the typical quantum interpretational problems.

Let us start by considering the main reason why a quantum entity cannot be considered a particle, then show that the same argument applies, \emph{mutatis mutandis}, to the field concept. A particle (or corpuscle) is, by definition, a system \emph{localized in space}. This means that if a physical entity is a particle then, in every moment, it must be characterizable by a specific position (for instance of its center of mass) in our three-dimensional Euclidean space. Let us call this fundamental attribute \emph{spatiality}.

Then, if \emph{microscopic} entities are assumed to obey \emph{Heisenberg's uncertainty principle} (HUP), as we know they do, one is forced to admit that the concept of ``microscopic particle'' is a self-contradictory one. This because if an entity obeys HUP, one cannot simultaneously determine its position and momentum and, as a consequence, one cannot determine, \emph{not even in principle}, how the position of the entity will vary in time (by solving the classical equations of motion). Consequently, one cannot \emph{predict with certainty} its future locations. 

Now, according to the \emph{reality criterion} formulated by Einstein, Podolsky and Rosen~\cite{Ein}, and further refined by Piron and Aerts~\cite{Pir,Aerts0,Sas1}, the notion of actual existence is intimately related to the notion of predictability, in the sense that a property can be said to be \emph{actual}, for a given physical entity, if and only if should one decide to observe it (i.e., to test it), the success of the observation would be in principle predictable in advance, with certainty.   

According to this general reality (or existence) criterion, one must conclude that a microscopic entity obeying the HUP cannot actually possess the property of being always present somewhere in space, as there are no means to predict its spatial localizations with certainty, not even in principle. Therefore, whatever its nature is, it is a \emph{non-spatial} entity, and if only for this reason it cannot be considered a particle~\cite{Sas1}. But then, if quantum entities are not particles, as they are intrinsically non-spatial, what can we say about their nature? Can we affirm, as suggested by Hobson, that they are mere fields, i.e., that the wave function $\psi_t$ has to be simply interpreted as a real \emph{space-filling} extended field?

Our point is that this cannot be done. Indeed, although classical fields, contrary to classical particles, are spatially extended entities, spread out over space, they are still spatial entities. Therefore, a field is an equally strongly spatial entity as a particle is. The only difference is that a particle is imagined to possess, at any moment of time, a specific, almost point-like location, whereas a field is imagined to have, at any moment of time, a spread out location in space. But a field is also an entity defined in space, possessing specific \emph{actual} properties in every point of it (like for instance force vectors). 

Thus, also the classical notion of field is unable to conveniently describe the non-spatial nature of a quantum entity. This becomes even more evident if one considers the situation of several quantum entities. Because of the experimentally verified presence of entanglement, the so-called fields of several quantum entities are certainly not fields that can be defined in a three-dimensional space, but only in a higher dimensional configuration space. 

Another way of showing that the wave function $\psi_t$ cannot describe an actual spatial field, is to study the notion of \emph{quantum sojourn time}, which measures the \emph{total availability} of a quantum entity in participating to a process of creation of a spatial localization~\cite{Sas2, Sas2bis}. Indeed, apart from the special case of bound states, one can show that the overall time potentially spent, on average, by a quantum entity in a given volume of space, say a ball of radius $r$ (defined as the integral $\int_{-\infty}^{\infty}dt \|P_r\psi_t\|^2$, where $P_r$ is the projection operator onto the set of states localized in the ball), is finite for all values of $r<\infty$. This is in contradiction with the hypothesis that $\psi_t$ would describe a space-filling extended field, permanently present in space, as if this would be the case then, clearly, such an average total time should be infinite~\cite{Sas2}.  

It is worth emphasizing that the possibility of understanding quantum non-spatiality is intimately related to the possibility of solving the measurement problem, at least at a conceptual level. Hobson rightly observes that the replacement of the concept of particle by the one of field cannot solve such a problem, but what we think he fails to consider is that both the particle and field concepts are inadequate because of the measurement issue.

A quantum ``field'' can certainly be understood in the abstract sense of a ``field of potentialities,'' i.e., a field of potential properties that can possibly be actualized -- in our three-dimensional space -- through measurement processes, i.e., through interactions with measuring apparatus, which are macroscopic entities stably present in our space. This transition from a ``potential mode of being'' to an ``actual mode of being,'' where each time only one among countless different possibilities is selected, is at the heart of a quantum measurement, and certainly needs to be considered if one wants to clarify the true nature of quantum entities. In ultimate analysis, what quantum mechanics teaches us is that not all of physical reality is contained within space, and that we need to drop the preconception that so-called microscopic ``particles'' and quantum ``fields'' would necessarily be spatial entities.~\cite{Aerts1,Aerts2,Sas1,Sas2,Sas3}.

\end{document}